\documentclass[aps,showpacs,prl,twocolumn,preprintnumbers]{revtex4}
\usepackage{epsfig}
\usepackage{amsmath,amsfonts,bm}
\begin{document}

\count255=\time\divide\count255 by 60 \xdef\hourmin{\number\count255}
  \multiply\count255 by-60\advance\count255 by\time
 \xdef\hourmin{\hourmin:\ifnum\count255<10 0\fi\the\count255}

\newcommand{\Dslash}{D\hspace{-0.7em}{ }\slash\hspace{0.2em}}
\newcommand\<{\langle}
\renewcommand\>{\rangle}
\renewcommand\d{\partial}
\newcommand\LambdaQCD{\Lambda_{\textrm{QCD}}}
\newcommand\tr{\mathop{\mathrm{Tr}}}
\newcommand\+{\dagger}
\newcommand\g{g_5}

\newcommand{\xbf}[1]{\mbox{\boldmath $ #1 $}}

\title {Production of  the Smallest QED Atom: True Muonium
($\mu^+ \mu^-$)}

\author{Stanley J. Brodsky}
\email{sjbth@slac.stanford.edu}

\affiliation{SLAC National Accelerator Laboratory, Stanford
University, Stanford, CA 94309, USA}

\author{Richard F. Lebed}
\email{Richard.Lebed@asu.edu}

\affiliation{Department of Physics, Arizona State University, Tempe,
AZ 85287-1504, USA}

\date{April, 2009}

\begin{abstract}
The ``true muonium" $(\mu^+ \mu^-)$ and ``true tauonium" $(\tau^+
\tau^-)$ bound states are not only the heaviest, but also the most
compact pure QED systems.  The rapid weak decay of the $\tau$ makes
the observation of true tauonium difficult.  However, as we show, the
production and study of true muonium is possible at modern
electron-positron colliders.
\end{abstract}

\preprint{SLAC-PUB-13575}

%

\maketitle

The possibility of a $\mu^+ \mu^-$ bound state, denoted here as
$(\mu^+ \mu^-)$, was surely realized not long after the
clarification~\cite{muon} of the leptonic nature of the muon, since
the first positronium calculations~\cite{Pirenne} and its
observation~\cite{Deutsch:1951zza} occurred in the same era.  The term
``muonium'' for the $\mu^+ e^-$ bound state and its first theoretical
discussion appeared in Ref.~~\cite{Friedman:1957mz}, and the state was
discovered soon thereafter~\cite{Hughes:1960zz}.  However, the first
detailed studies~\cite{Hughesmumu,Bilenkii} of $(\mu^+ \mu^-)$
(alternately dubbed ``true muonium''~\cite{Hughesmumu} and
``dimuonium''~\cite{Malenfant:1987tm,Karshenboim:1998we}) only began
as experimental advances made its production tenable.  Positronium,
muonium, $\pi \mu$ atoms~\cite{Coombes:1976hi}, and more recently even
dipositronium [the $(e^+ e^-)(e^+ e^-)$ molecule]~\cite{dipos} have
been produced and studied, but true muonium has not yet been produced.

\begin{table}
\caption{True fermionium decay times and their ratios.}
\label{tauformulae}
\begin{tabular}{ll}
\hline\hline \\
$\tau (n {}^3S_1 \to e^+ e^-) =
\frac{\textstyle 6\hbar n^3}{\textstyle \alpha^5 mc^2}$ , &
$\tau (n {}^1S_0 \to \gamma \gamma) =
\frac{\textstyle 2\hbar n^3}{\textstyle \alpha^5 mc^2}$ ,
\vspace{2ex}\\
$\tau (2P \! \to \! 1S) = \left( \frac{\textstyle 3}{\textstyle 2}
\right)^8 \! \! \frac{\textstyle 2\hbar}{\textstyle \alpha^5 mc^2}$ ,
& 
$\tau (3S \! \to \! 2P) = \left( \frac{\textstyle 5}{\textstyle 2}
\right)^9 \! \! \frac{\textstyle 4\hbar}{\textstyle 3\alpha^5 mc^2}$ ,
\vspace{2ex}
\end{tabular}
\begin{tabular}{ll}
$\frac{\textstyle \tau (n {}^3S_1 \to e^+ e^-)}
{\textstyle \tau(n {}^1S_0 \to \gamma \gamma)} = 3$ , &
$\frac{\textstyle \tau (2P \to 1S)}
{\textstyle \tau (n {}^1S_0 \to \gamma \gamma)} = \left(
\frac{\textstyle 3}{\textstyle 2} \right)^8 \! \! \frac{\textstyle
1}{\textstyle n^3} = \frac{\textstyle 25.6}{\textstyle n^3}$ ,
\vspace{2ex}
\end{tabular}
\begin{tabular}{c}
$\frac{\textstyle \tau (3S \to 2P)}{\textstyle \tau (2P \to 1S)}
= \left( \frac{\textstyle 5}{\textstyle 3} \right)^9 = 99.2$ .
\end{tabular}
\end{table}
\begin{figure}
\hspace{-7.4em}
\includegraphics[scale=0.29]{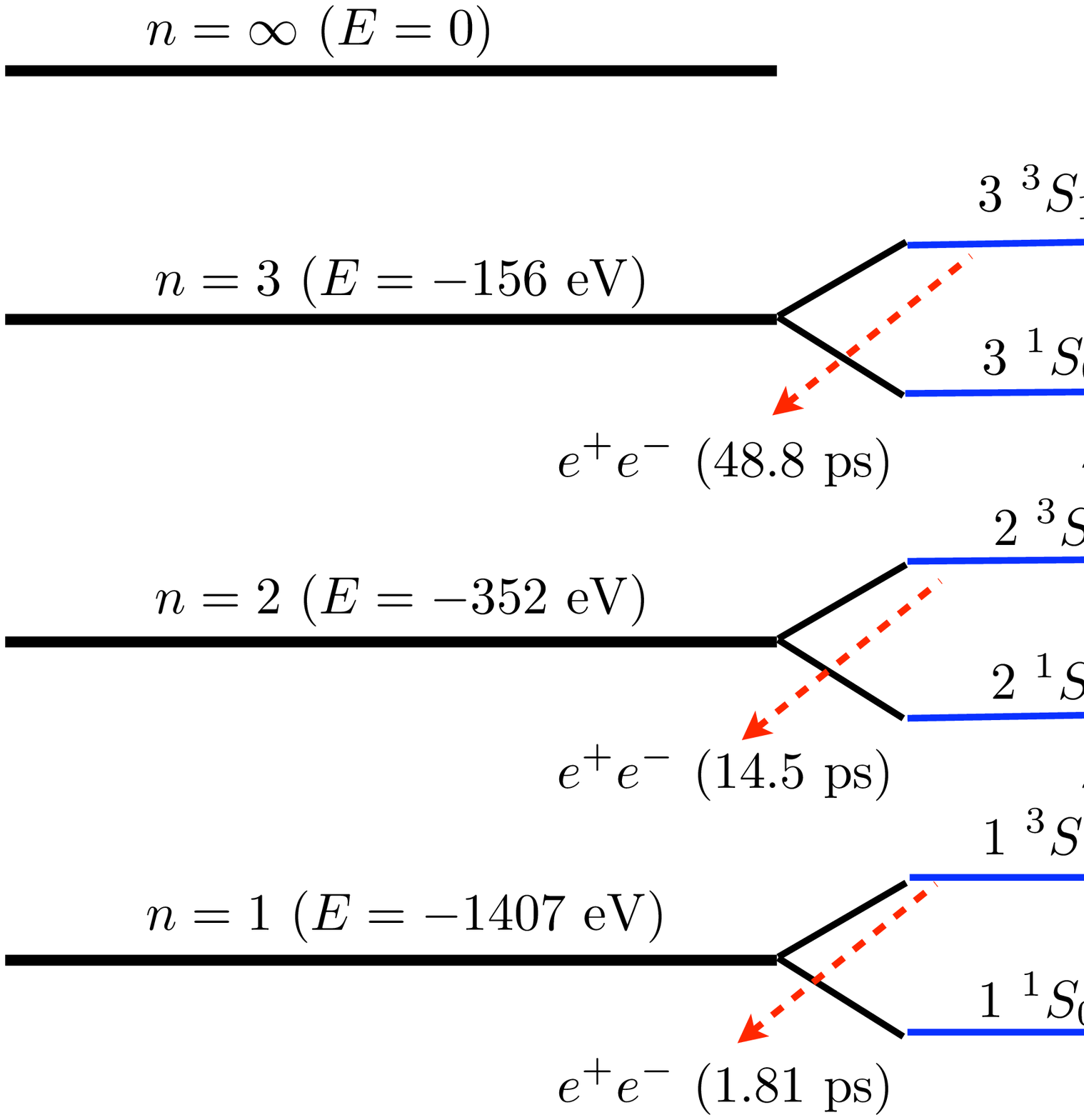}
\caption{\label{level} True muonium level diagram (spacings not to
scale).}
\end{figure}
The true muonium $(\mu^+ \mu^-)$ and true tauonium $(\tau^+ \tau^-)$
[and the much more difficult to produce ``mu-tauonium'' $(\mu^\pm
\tau^\mp)$] bound states are not only the heaviest, but also the most
compact pure QED systems [the $(\mu^+ \mu^-)$ Bohr radius is 512~fm].
The relatively rapid weak decay of the $\tau$ unfortunately makes the
observation and study of true tauonium more difficult, as quantified
below.  In the case of true muonium the proposed production mechanisms
include $\pi^- p \! \to \! (\mu^+ \mu^-) n$~\cite{Bilenkii}, $\gamma Z
\! \to \!  (\mu^+ \mu^-) Z$~\cite{Bilenkii}, $e Z \! \to \! e (\mu^+
\mu^-) Z$~\cite{Holvik:1986ty}, $Z_1 Z_2 \! \to \! Z_1 Z_2 (\mu^+
\mu^-)$~\cite{Ginzburg:1998df} (where $Z$ indicates a heavy nucleus),
direct $\mu^+ \mu^-$ collisions~\cite{Hughesmumu}, $\eta \! \to \!
(\mu^+ \mu^-) \gamma$~\cite{etadecay}, and $e^+ e^- \! \to \! (\mu^+
\mu^-)$~\cite{Moffat:1975uw}.  In addition, the properties of true
muonium have been studied in a number of
papers~\cite{OR,Karshenboim:1998we,Jentschura}.

The $e^+ e^- \! \to \! (\mu^+ \mu^-)$ production mechanism is
particularly interesting because it contains no hadrons, whose
concomitant decays would need to be disentangled in the reconstruction
process.  If the beam energies of the collider are set near threshold
$\sqrt s \! \sim \! 2 m_\mu, $ the typical beam spread is so large
compared to bound-state energy level spacings that every $nS$ state is
produced, with relative probability $\sim 1/n^3$ [{\it i.e.}, scaling
with the $(\mu^+ \mu^-)$ squared wave functions $|\psi_{n00}(0)|^2$ at
the interaction point, $r \! = \! 0$] and carrying the Bohr binding
energy $-m_\mu \alpha^2/4n^2.$ The high-$n$ states are so densely
spaced that the total cross section is indistinguishable~\cite{Bj}
from the rate just above threshold, after including the
Sommerfeld-Schwinger-Sakharov (SSS) threshold enhancement
factor~\cite{SSSfactor} from Coulomb rescattering.  As discussed below
[Eq.~(\ref{SSSeq}) and following], the SSS correction $\sim \pi \alpha
/ \beta$ cancels the factor of $\beta$, the velocity of either of
$\mu^{\pm}$ in their common center-of-momentum (c.m.) frame, that
arises from phase space.

The spectrum and decay channels for true muonium are summarized in
Fig.~\ref{level}, using well-known quantum mechanical
expressions~\cite{Mizushima} collected in Table~\ref{tauformulae}.  In
most cases, the spectrum and decay widths of $(\mu^+ \mu^-)$ mimic the
spectrum of positronium scaled by the mass ratio $m_\mu /m_e$.
However, while positronium of course has no $e^+ e^-$ decay channels,
$(\mu^+ \mu^-)[n{}^3S_1] \! \to \! \gamma^* \! \to \! e^+ e^-$ is
allowed and has a rate and precision spectroscopy sensitive to vacuum
polarization corrections via the timelike running coupling $\alpha
(q^2 \! > \! 0)$.

Unlike the case of positronium, the ($\mu^+ \mu^-$) constituents
themselves are unstable.  However, the $\mu$ has an exceptionally long
lifetime by particle physics standards (2.2~$\mu {\rm s}$), meaning
that ($\mu^+ \mu^-$) annihilates long before its constituents weakly
decay, and thus true muonium is unique as the heaviest metastable
laboratory possible for precision QED tests: $(\mu^+ \mu^-)$ has a
lifetime of $0.602$~ps in the ${}^1S_0$ state (decaying to $\gamma
\gamma$)~\cite{Bilenkii,Hughesmumu} and $1.81$~ps in the ${}^3S_1$
state (decaying to $e^+ e^-$)~\cite{Bilenkii}.

In principle, the creation of true tauonium ($\tau^+ \tau^-$) is also
possible; the corresponding ${}^1S_0$ and ${}^3S_1$ lifetimes are
35.8~fs and 107~fs, respectively, to be compared with the free $\tau$
lifetime 291~fs (or half this for a system of two $\tau$'s).  One sees
that the ($\tau^+ \tau^-$) annihilation decay and the weak decay of
the constituent $\tau$'s actually compete, making $(\tau^+ \tau^-)$
not a pure QED system like $(e^+ e^-)$.

Electron-positron colliders have reached exceptional luminosity
values, leading to the possibility of detecting processes with very
small branching fractions.  The original proposal by
Moffat~\cite{Moffat:1975uw} suggested searching for X-rays from
($\mu^+ \mu^-)$ Bohr transitions such as $2P\! \to \! 1S$ at
directions normal to the beam.  However, the $nS$ states typically
decay via annihilation to $e^+ e^-$ and $\gamma \gamma$ before they
can populate longer-lived states.  Furthermore, the production and
rapid decay of a single neutral system at rest or moving in the beam
line would be difficult to detect relative to the continuum QED
backgrounds, due to a preponderance of noninteracting beam particles
and synchrotron radiation.

In this letter we propose two distinct methods for producing a moving
true muonium atom in $e^+ e^-$ collisions.  In both methods the motion
of the atom allows one to observe a gap between the production point
at the beam crossing and its decay to $e^+ e^- $ or $\gamma \gamma$
final states.  Furthermore, each given lifetime is enhanced by a
relativistic dilation factor $\gamma$ appropriate to the process.

\begin{figure}
\includegraphics[scale=0.8]{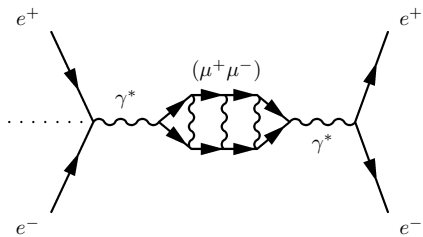}
\caption{\label{FISR} The ``Fool's ISR'' configuration for
$e^+ e^- \! \to (\mu^+ \mu^-)$ for symmetric beam energies.  The angle
between the either of the $e^\pm$ and dotted line ($\hat z$ axis) is
defined as $\theta$.}
\end{figure}
In the first method, we utilize an $e^+ e^-$ collider in which the
atomic system produced in $e^+ e^- \! \to \gamma^* \! \to \! (\mu^+
\mu^-)$ at $s \! \simeq \! 4m^2_\mu$ carries momentum $\vec p = \vec
p_{e^+} \! + \vec p_{e^-} \! \neq \! 0$.  The production point of the
($\mu^+ \mu^-$) and its decay point are thus spatially displaced along
the beam direction.
Asymmetric $e^+ e^-$ colliders PEP-II and KEKB have been utilized for
the BaBar and Belle experiments.  However, we propose configuring an
$e^+ e^-$ collider to use the ``Fool's Intersection Storage Ring''
(FISR) discussed by Bjorken~\cite{Bjorken:1976mk} (Fig.~\ref{FISR}) in
which the $e^\pm$ beams are arranged to merge at a small angle
$2\theta$ (bisected by $\hat z$), so that $s \! = \! (p_{e^+} \!  +
p_{e^-} \!)^2 \simeq 2 E_+ E_- (1 - \cos 2\theta) \simeq 4 m^2_\mu$
and the atom moves with momentum $p_z = (E_+ + E_-) \cos \theta.$ For
example, for $\theta \! = \! 5^\circ$ and equal-energy $e^{\pm}$ beams
$E_\pm = 1.212$~GeV, the atom has lab-frame momentum $p_z = 2.415 $
GeV and $\gamma = E_{\rm lab} / 2 m_\mu = 11.5$.  One can thus utilize
symmetric or asymmetric beams in the GeV range colliding at small
angles to obtain the c.m.\ energy $ \sqrt s \simeq 2 m_\mu$ for the
production of true muonium.

The gap between the formation of the atom and its decay as it
propagates should be clearly detectable since its path lies in neither
beam pipe.  The $3{}^3S_1$ state decays with a $50$~ps lifetime, so it
moves $1.5$~cm of proper distance before decaying to $e^+ e^-$, a
length enhanced in the lab frame by the $\gamma$ factor (to 16.8~cm in
the $\theta \! = \! 5^\circ$ example).  One thus can observe the
appearance of $e^+ e^-$ events with a $\theta$-dependent set of
lifetimes.

The cross section for continuum muon pair production $e^+ e^-
\! \to \! \mu^+ \mu^-$ just above threshold is the Born cross section
enhanced by the Sommerfeld-Schwinger-Sakharov (SSS)
threshold Coulomb resummation factor~\cite{SSSfactor} $S(\beta)$:
\begin{equation}
\sigma = \frac{2 \pi \alpha^2 \beta}{s} \left( 1 - \frac{\beta^2}{3}
\right) S(\beta) \, ,
\end{equation}
where 
\begin{equation} \label{SSSeq}
S(\beta) = \frac{X(\beta)}{1-\exp[-X(\beta)]} \, .
\end{equation}
Here $\beta = \sqrt{1- 4m^2_\mu/s}$ is the velocity of either of the
$\mu^\pm$ in their c.m.\ frame, and $X(\beta) \! = \! \pi \alpha
\sqrt{1-\beta^2}/\beta$.  Thus the factor of $\beta$ due to phase
space is cancelled by the SSS factor, so that continuum production
occurs even at threshold where $\beta \! = \! 0$.
For values of $|\beta|$ of order $\alpha$ (as in Bohr bound states),
we see that the SSS factor effectively replaces $\beta$ with $\pi
\alpha$.  Below threshold the entire set of ortho-true muonium
$n^3S_1$ $C \! = \! -1$ Bohr bound-state resonances with $n = 1, 2,
\cdots$ is produced, with weights $\sim \! 1/n^3$ and spaced
with increasing density according to the Bohr energies $(\sqrt s)_n
\simeq 2 m_\mu \! - \! {\alpha^2 m_\mu/4 n^2}$.  By duality, the rates
smeared over energies above and below threshold should be
indistinguishable~\cite{Bj}.  Thus the total production of bound
states in $e^+ e^- \! \to \! (\mu^+ \mu^-)$ relative to the $e^+ e^-
\! \to \! \mu^+ \mu^-$ relativistic lepton pair rate is of order $R
\! \sim \! \frac 3 2 \pi \alpha \! \simeq \! 0.03$.  However, in
practice the production rate is also reduced by the Bohr energy
divided by the finite width of the beam energies, since only
collisions in the energy window $\delta E \simeq m_\mu \alpha^2/4$ are
effective in producing bound states.  For example, if the beam
resolution is of order $\Delta E_\pm = 0.01~m_\mu \! \sim \! 1$~MeV,
the effective $R$ is reduced by a further ${\delta E/ \Delta E_\pm}
\simeq 10^{-3}$, leading to a production of $(\mu^+ \mu^-)$ at
$\sim \! 5 \! \times \! 10^{-5}$ of the standard $e^+ e^- \! \to
\ell^+ \ell^-$ rate.

Since the $(\mu^+ \mu^-)$ state in the FISR method is produced through
a single $C \! = \! -$ photon, $n^3S_1$ states (and not $n^1S_0$) are
produced, which decay almost always to $e^\pm$ pairs, as illustrated
in Fig.~\ref{FISR}.  Note that this holds even for radiative
transitions through the sequences $n^{\prime \prime}{}^3S_1 \to
n^\prime P \to n{}^3S_1$, since the intermediate $P$ states do not
annihilate.

The $(\mu^+ \mu^-)$ bound states, once produced, can in principle be
studied by exposure to $O({\rm ps})$ laser or microwave bursts, or
dissociated into free $\mu^\pm$ by passing through a foil.  Because of
the novel kinematics of the FISR, the true muonium state can be
produced within a laser cavity.  For example, an intense laser source
can conceivably excite the $nS$ state of the atom to a $P$ state
before the former's annihilation decay.  A $2P$ state produced in this
way has a lifetime of (15.4~ps)$\times \gamma$.  In principle this
allows precision spectroscopy of true muonium, including measurements
of the $2P$-$2S$ and other splittings.  Laser spectroscopy of $(\mu
Z)$ atoms is reviewed in Ref.~\cite{Jung}.

In this letter we also propose a second production mechanism, $e^+ e^-
\! \to \! \gamma (\mu^+ \mu^-)$, which can be used for high-energy
colliding beams with conventional configuration. It has the advantage
that the production rate is independent of beam resolution, and
removes the ($\mu^+ \mu^-$) completely from the beam line since the
atom recoils against a co-produced hard $\gamma$.  While the
production of the real $\gamma$ costs an additional factor of $\alpha$
in the rate, the kinematics is exceptionally clean: Since the process
is quasi-two-body, the $\gamma$ is nearly monochromatic [neglecting
the $(\mu^+ \mu^-)$ binding energy] as a function of the total c.m.\
beam energy $\sqrt{s}$, $E_\gamma \! = \! (s - 4m_\mu^2)/2\sqrt{s}$.
Furthermore, the $(\mu^+ \mu^-)$ lifetime is enhanced by the dilation
factor $\gamma \!  = \!  (s + 4m_\mu^2)/4m_\mu \sqrt{s}$.
\begin{figure}
\includegraphics[scale=0.7]{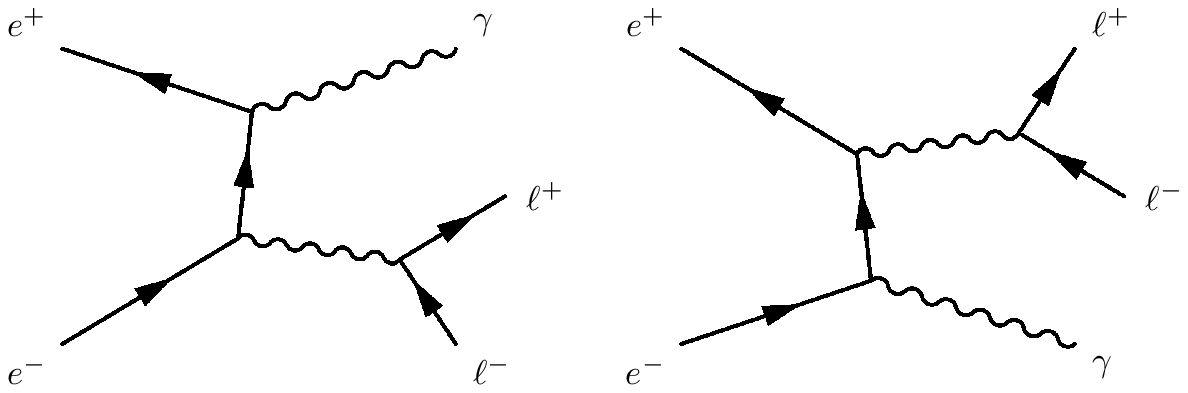}
\includegraphics[scale=0.7]{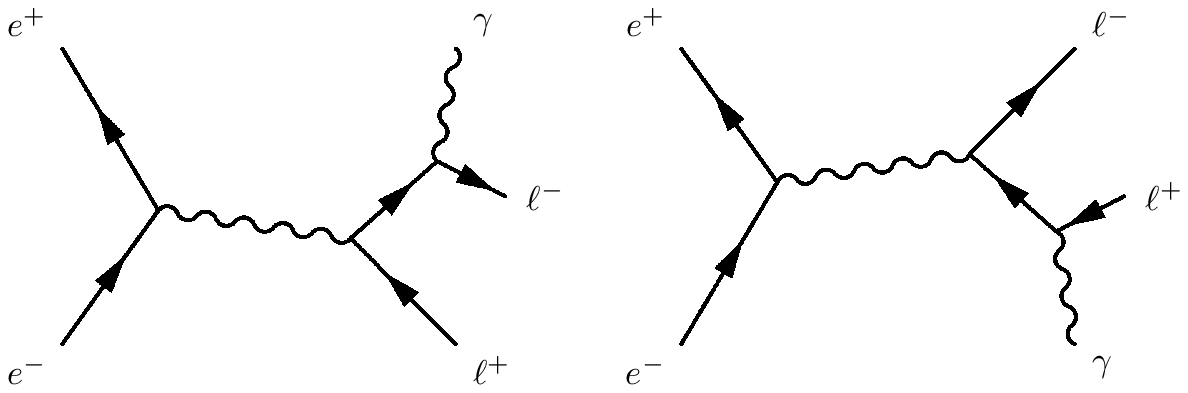}
\caption{\label{Feyn} Dominant Feynman diagrams for $e^+ e^- \!
\to \! \gamma (\ell^+ \ell^-)$.}
\end{figure}
The dominant Feynman diagrams for $e^+ e^- \! \to \! \gamma (\ell^+
\ell^-)$ are shown in Fig.~\ref{Feyn}.  If $E_\gamma$ is large
compared to $m_\mu$, and its angle from the beam is large, one also
can have significant radiation from the $\mu$ lines.  In fact, if the
hard $\gamma$ is emitted by a $\mu$, the true muonium state is formed
in the para $n^1S_0$ $C \! = \! +$ state, which leads to two-photon
annihilation decays.  In this case both $e^+ e^-$ and $\gamma \gamma$
final states appear, accompanied by a decay gap.

The $O(\alpha^3)$ Born amplitude for the process $e^+ e^- \! \to \!
\gamma \mu^+ \mu^-$ (free $\mu$'s) in the kinematic regime $m_e^2/s$,
$m_\mu^2/s \! \ll \! 1$ was first computed long ago in
Ref.~\cite{Kuraev:1977rp}.  More recently, a related
collaboration~\cite{Arbuzov:2004gy} specialized this calculation to
precisely the desired kinematics: The invariant mass square $s_1$ of
the $\mu^\pm$ pair is assumed small compared to the total c.m.\
squared energy $s$.  In this case, the Born differential cross section
for the diagrams in Fig.~\ref{Feyn} is
\begin{equation} \label{Kuraev_calc}
d\sigma = \frac{\alpha^3 (1+c^2)}{s s_1 (1-c^2)} \left( 2 \delta + 1 -
2 x_- x_+ \right) dx_- \, dc \, ds_1 \, ,
\end{equation}
where $c$ is defined as the cosine of the angle between the $e^-$ and
$\mu^-$ [and hence also the ($\mu^+ \mu^-$) atom], $\delta \! \equiv
\!  m_\mu^2/s_1 \! \simeq \! \frac 1 4$ for ($\mu^+ \mu^-$) bound
states, $x_\pm$ are the fractions of the half of the c.m.\ beam energy
$E_{\pm}/(\sqrt{s}/2)$ that is carried by $\mu^\pm$ (the other half
being carried by the $\gamma$) so that $x_+ \! + x_- \!  = \!  1$, and
the range of $x_\pm$ is given by
\begin{equation} \label{xmrange}
\frac 1 2 (1 - \beta^\prime) \leq x_{\pm} \leq
\frac 1 2 (1 + \beta^\prime) \, ,
\end{equation}
where $\beta^\prime \! \equiv \! \sqrt{1 \! - \! 4\delta}$ is the
velocity of either of the $\mu^\pm$ in their c.m.\ frame.  In
addition, the hard photon momentum makes an angle $\theta_\gamma$ with
the initial $e^-$ that is assumed to lie outside of narrow cones of
opening angle $\theta_0$ around the beam axis, $\theta_0 \! < \!
\theta_\gamma \! < \! \pi \! - \! \theta_0$, where $2m_\mu /\sqrt{s}
\! \ll \! \theta_0 \! \ll \! 1$.  Note that the $\gamma$ and ($\mu^+
\mu^-$) are back-to-back, $\theta_\gamma \! = \! \pi \! - \! \theta$.

The differential cross section in Eq.~(\ref{Kuraev_calc}) becomes
singular when the $\gamma$ [and hence also the ($\mu^+ \mu^-$)] is
collinear with the beam.  For the purpose of our cross section
estimates, we integrate $c$ over the range excluding the beam cone, $c
\! \in \! [-c_0, c_0]$, where $c_0 \! \equiv \! \cos \theta_0$.
Using also Eq.~(\ref{xmrange}) to integrate over $x_-$, one obtains
\begin{equation} \label{Kuraev_int}
\frac{d\sigma}{ds_1} = 2 \beta^\prime \left[ \ln
\left( \frac{1+c_0}{1-c_0} \right) - c_0 \right]
\frac{\alpha^3}{ss_1} \, .
\end{equation}
The factor $\beta^\prime$, indicating that the cross section vanishes
at $\mu^\pm$ threshold, arises simply from 3-body phase space.

Equations~(\ref{Kuraev_calc}) and (\ref{Kuraev_int}) describe a
process in which the $\mu^\pm$ pair carry an invariant mass $s_1$
small compared to $s$, but are not necessary bound together.  In order
to compute the cross section for such a process, one must again
include the SSS threshold Coulomb resummation factor~\cite{SSSfactor}.
Here the $\beta^\prime$ in Eq.~(\ref{Kuraev_int}) refers to the
(continuous) velocity of each of a free $\mu^\pm$ pair in its c.m.\
frame, whereas $\beta^\prime$ in the bound-state formalism refers to
the (quantized) velocity of each particle within their Coulomb
potential well.  Nevertheless, as argued in the previous case, by
duality the same cross section formulae still hold in the bound-state
regime if the SSS factor is taken into account and the weights of the
discrete transitions are properly included and smeared over the
allowed energy range for bound states~\cite{Bj}.  One then obtains
\begin{equation}
\frac{d\sigma}{ds_1} = 2 \pi \left[ \ln \left( \frac{1+c_0}{1-c_0}
\right) - c_0 \right] \frac{\alpha^4}{ss_1} \, .
\end{equation}
The relevant range of $ds_1$ is just that where bound Bohr states
occur, which begin
at energy $\alpha^2 m_\mu / 4$ below the pair creation threshold $s_1
\! = \! 4m_\mu^2$, and thus give rise to $ds_1 \! \simeq \!  m_\mu^2
\alpha^2$.  Thus one obtains
\begin{equation}
\sigma \simeq \frac{\pi}{2} \left[ \ln \left( \frac{1+c_0}{1-c_0}
\right) - c_0 \right] \frac{\alpha^6}{s} \, .
\end{equation}
The angular factor is again singular for $c_0 \! = \! \pm 1$, varying
from zero at $\theta_0 \! = \! \pi/2$, to unity near $\pi/4$, to over
7 at 2$^\circ$.

Note that $\beta \! \equiv \! \sqrt{1 \! - \! 4m_\mu^2/s}$ differs
from $\beta^\prime$ used for $e^+ e^- \! \to \! \gamma (\mu^+ \mu^-)$;
for processes $e^+ e^- \! \to \! \gamma (\mu^+ \mu^-)$ with the same
value of $s$ for which Eq.~(\ref{Kuraev_calc}) and following are
applicable, recall that $m_\mu^2 \! \ll \! s$ and hence $\beta \!
\simeq \! 1$.  The ratio $\sigma [e^+ e^- \! \to \! \gamma (\mu^+
\mu^-)] / \sigma (e^+ e^- \! \to \! \mu^+ \mu^-)$ at the same value of
$s$ is therefore just a number close to unity ({\it e.g.}, 2.66 for
$\theta_0 \! = \!  2$~degrees) times $\alpha^4$.  While this
$O(10^{-8})$ suppression may seem overwhelming, it is within the
capabilities of modern $e^+ e^-$ facilities.  For example, the BEPCII
peak luminosity will be $10^{33}
\, {\rm cm}^{-2} {\rm s}^{-1}$ at a c.m.\ energy of 3.78~GeV, but
varying between 2 and 4.6~GeV~\cite{Harris:2008tx}.  At 2~GeV about 5
events $e^+ e^- \! \to \! \gamma (\mu^+ \mu^-)$ occur per year of run
time, and the yield increases with $1/s$.  On the other hand, for
smaller values of $s$ the dilation factor $\gamma$ for $(\mu^+ \mu^-)$
becomes shorter, thus diminishing its lifetime and hence track length.

The production is much more prominent if one performs a cut on $s_1$
values near the $\mu^\pm$ threshold $4m_\mu^2$.  In that case one
should compare Eq.~(\ref{Kuraev_int}) to the derivative $d\sigma (e^+
e^- \! \to \! \mu^+ \mu^-)/ds$ [which is not actually a differential
cross section but rather the difference of $\sigma (e^+ e^- \! \to \!
\mu^+ \mu^-)$ between bins at c.m.\ squared energies $s$ and $s \! +
\!  ds$].  Then the relative suppression is only $O(\alpha^2)$, one
$\alpha$ arising from the extra photon and one from the SSS factor.

Between the two proposals presented here, $e^+ e^- \! \to \! (\mu^+
\mu^-)$ with beams merging at a small crossing angle, and the rarer 
process $e^+ e^- \! \to \gamma (\mu^+ \mu^-)$ that can access both
ortho and para states with conventional beam kinematics, the discovery
and observation of the true muonium atom $(\mu^+ \mu^-)$ appears to be
well within current experimental capabilities.

\vspace{-0.5em}
\section*{Acknowledgments}
\vspace{-0.5em}
This research was supported under DOE Contract DE-AC02-76SF00515 (SJB)
and NSF Grant No.\ PHY-0757394 (RFL).  SJB thanks Spencer Klein and
Mike Woods for useful conversations, and RFL thanks the SLAC Theory
Group, where this work was inspired, for their hospitality.

\end{document}